# Multi-Tenant Radio Access Network Slicing: Statistical Multiplexing of Spatial Loads

P. Caballero, A. Banchs, *Senior Member, IEEE,* G. de Veciana, *Fellow, IEEE,* and X. Costa-Pérez, *Member, IEEE*

*Abstract*—This paper addresses the slicing of Radio Access Network (RAN) resources by multiple tenants, e.g., virtual wireless operators and service providers. We consider a criterion for dynamic resource allocation amongst tenants, based on a weighted proportionally fair objective, which achieves desirable fairness/protection across the network slices of the different tenants and their associated users. Several key properties are established, including: the Pareto-optimality of user association to base stations, the fair allocation of base stations' resources, and the gains resulting from dynamic resource sharing across slices, both in terms of utility gains and capacity savings. We then address algorithmic and practical challenges in realizing the proposed criterion. We show that the objective is NP-hard, making an exact solution impractical, and design a distributed semi-online algorithm which meets performance guarantees in equilibrium and can be shown to quickly converge to a region around the equilibrium point. Building on this algorithm, we devise a practical approach with limited computational, information, and handoff overheads. We use detailed simulations to show that our approach is indeed near-optimal and provides substantial gains both to tenants (in terms of capacity savings) and end-users (in terms of improved performance).

## I. Introduction

Driven by the capacity requirements forecasted for future mobile networks as well as the decreasing margins obtained by operators, infrastructure sharing has established itself as a key business model for mobile operators to reduce the deployment and operational costs of their networks (e.g., [1] reports a 280% increase in deals within the last 5 years). While passive and active sharing solutions, ranging from exclusive allocation of resources to roaming agreements, are used and have been standardized, these sharing approaches are based on fixed contractual agreements with Mobile Virtual Network Operators (MVNO) over long time periods (typically on a monthly/yearly basis). In this paper, we focus on a structured dynamic slicing approach which enables a much more efficient sharing of network resources, as envisioned by the 3GPP Network Sharing Enhancements for future mobile networks which the authors contributed to [2]. Following [3], our approach divides the infrastructure into *network slices*, assigning a different slice to each operator, and implements the sharing of network resources among operators by dynamically allocating resources to slices.

Such a novel network slicing approach is expected to result in new business models and revenue sources for infrastructure providers.[1] Indeed, this approach supports not only classical players (mobile operators) but also new ones such as Over-The-Top (OTT) service providers that may buy a *slice* of the network to ensure satisfactory service to their users (e.g., Amazon Kindle's support for downloading content or a pay TV channel including a premium subscription). In the literature, the term *tenants* is often used to refer to the different types of players, and *multi-tenancy* refers to approaches enabling dynamic network slicing and resource sharing for multiple tenants. For simplicity, hereafter we use the term operator in a broad sense to refer to classical (virtual) operators as well as the new players enabled by this approach.

In designing a practical solution for dynamic resource sharing among slices we face multiple challenges. To start with, we need a *sharing criterion* that not only allocates resources to operators (and their corresponding slices) fairly, but also, shares the resources of each operator fairly among its users. Furthermore, the criterion should allow for flexible sharing "levels" to meet operators' heterogeneous requirements; for practical purposes, these levels should be coarse-grained, rather than based on instantaneous resource needs. When allocating resources to an operator, one should take into account the numbers and locations of active users on the network – indeed some locations may see higher demand and (consequently) the associated resources might be scarce.

Beyond the criterion itself, designing an algorithm to implement it, while realizing timely adaptation to network changes, is also very challenging. Given the amount of information involved (including the channel quality of each user) and its dynamic nature, the algorithm should be as *distributed* as possible. Also, since the algorithm may be triggered frequently (whenever a user joins, leaves or changes its location), it should be *computationally efficient*. When adapting to network changes, the algorithm should control the number of handoffs triggered, as those may represent a high *overhead*.

*Key contributions*: This paper proposes a criterion for slicing the network infrastructure amongst operators and an algorithm to allocate resources accordingly. The key contributions are as follows. In Section II, we introduce a criterion for dynamic resource sharing among operators; while the criterion has been proposed before, we provide a characterization supporting its use in a multi-tenant network setting. These properties are developed in Section II-C, providing insights on the optimality and fairness of the resulting allocations, and the benefits are studied in Section II-D, by characterizing the capacity savings by means of a closed-formula. We show that the criterion not only improves overall network utility but also that of each individual operator, thus guaranteeing that operators are not harmed by the sharing of resources amongst slices.

P. Caballero and G. de Veciana are with the University of Texas at Austin. A. Banchs is with the University Carlos III of Madrid and with the Institute IMDEA Networks, Spain. X. Costa-Pérez is with NEC Europte Ltd., Germany.

[1] See for instance the envisioned roles defined by the organization of telecommunication operators NGMN for 5G [3].

In Section III-A, we show the criterion corresponds to an NP-hard problem, motivating the need to devise an efficient approximation algorithm which is introduced in Section III-B. The proposed algorithm is semi-online, distributed, incurs low computational complexity, and has been specifically designed to control overheads associated with handoffs and/or mobile user reassociations. We rely on several intermediate analytical results to drive the key design choices underlying our algorithm. One of these intermediate results, presented in Section III-B2, is a variation of the algorithm which achieves similar performance bounds to state-of-the art algorithms while being distributed – which is in itself a valuable theoretical contribution on state-of-the-art approaches. Section IV provides a comprehensive performance evaluation based on detailed simulations, showing that ($i$) operators can save up to 80% capacity while providing the same quality to their users, and ($ii$) for a fixed capacity, we improve user performance in terms of file download times by up to 30%, among other results.

*Related work*: We next review and contrast our work to the state-of-the-art in ($i$) resource allocation based on proportional fairness, and ($ii$) resource sharing among operators.

Considerable research effort has been devoted to address the problem of fair resource allocation in networks. In wireline networks, fair resource allocation based on utility function maximization has been extensively studied following the seminal work of [4]. Building on this work, further algorithms for congestion control in multi-path environments have been proposed [5]–[7]. Not unlike our work, these algorithms are distributed. However, they allow users to decide among multiple routes while we focus on a wireless setting where each user can only use one resource (her base station).

In the specific context of wireless networks, several approaches have been proposed [8]–[10] to the problem of resource allocation and user association based on weighted and unweighted proportional fairness, respectively. The unweighted case has been largely studied in the literature in different contexts (e.g., power control [11], interference avoidance [12]). The authors of [8] and [12] analyzed the complexity of the problem and proved the existence of polynomial time algorithms which provide an exact solution, and [10] designed a distributed algorithm with convergence guarantees. In contrast to the above, the resource allocation criterion proposed in this paper relies on *weighted* proportional fairness, with operator-specific weights; this is a more difficult problem as it is NP-hard [8] and the convergence of distributed greedy algorithms cannot be guaranteed [13].

Weighted proportional fairness resource allocation in wireless networks has received much less attention. In [9], an algorithm with tight worst-case performance bounds of $\log(2)$ is proposed, while [14] proposes an heuristic algorithm. In contrast to the distributed approach proposed in this paper, both algorithms are centralized, which are expected to perform better since they possess global network information. However, the performance bound of the proposed distributed algorithm is $\log(e)$, close to the bound in [9]. The authors of [15], [16] propose a Gibbs-sampling mechanism based on simulated annealing that converge to an optimal solution. However, the convergence of this mechanism is known to be slow and for this reason they resort to a greedy solution, with the same spirit of our algorithm. For the latter, the authors neither provide performance bounds nor analyze convergence of the proposed algorithm; additionally, the overhead of these algorithms is not controlled, which limits their practical deployment. All the approaches mentioned above address the problem of a single-operator network, in contrast to our work which focuses on the slicing and sharing of resources among multiple operators.

Multi-operator network sharing has been studied from many different angles, including planning [17], economics [18], coverage [19], performance [20], etc. This paper focuses specifically on the design of algorithms for resource sharing among operators, which has been previously addressed by [21]–[23]. In [21], the optimization of the total network utility is addressed by using max-min fairness; in contrast, the criterion that we propose here relies on weighted proportional fairness, which (as we show) provides many desirable properties. The works of [22], [23] present a proportional fair formulation similar to ours; however, they do not provide a rationale for their choice, in contrast to the solid analytical arguments provided in this paper. Furthermore, [22] does not address the algorithm design, while [23] uses a general non-linear solver that incurs a very high computational complexity (as confirmed by our results of Section IV-D). To the best of our knowledge, this paper is the first to propose an efficient algorithm for multi-tenant network slicing that builds on analytical results.

## II. Resource Allocation Criterion

In this section, we formulate the optimization problem that will drive ($i$) the association of users to base stations, and ($ii$) the allocation of base stations' resources to users. Hereafter, we refer to this optimization as the *multi-operator resource allocation* (MORA) criterion. We show analytically that the criterion satisfies desirable properties in terms of optimality and fairness, and develop a simple model to evaluate the potential sharing gains of our network slicing approach.

### A. System model

We start by presenting our system model which was developed with LTE/LTE-A systems in mind, but is generally applicable to cellular systems. Consider a network consisting of a set $\mathcal{B}$ of base stations (or sectors in case of sector antennas) that are shared by a set of operators $\mathcal{O}$. At any given time, we let $\mathcal{U}$ denote the set of users sharing the network and $\mathcal{U}_o, o \in \mathcal{O}$ the subsets of users belonging to each operator. We let $c_{ub}$ denote the current transmission rate from base station $b$ to user $u$, which is defined as the rate the user would receive if she were allocated all of the base station's resources with the modulation-coding scheme selected for transmissions to the user based among others on path loss, neighbouring stations interference and fast fading; following similar analyses in the literature [21]–[25], we shall assume that $c_{ub}$ is fixed for each {user, base station} pair. An allocation of resources involves two sets of variables: ($i$) the association of users to base stations, denoted by $\mathbf{x} = (x_{ub} : u \in \mathcal{U}, b \in \mathcal{B})$, where each user $u$ is associated with a single base station, i.e.,

$x_{ub} = 1$ for one of the base stations and 0 otherwise, and (ii) the allocation of the resources of each base station among its associated users, denoted by $\mathbf{f} = (f_{ub} : u \in \mathcal{U}, b \in \mathcal{B})$, where $f_{ub}$ is the fraction of the base station $b$'s resources which are allocated to user $u$. Thus the rate allocated to user $u$, $r_u(\mathbf{x}, \mathbf{f})$, under a given allocation of resources, is given by

$$r_u(\mathbf{x}, \mathbf{f}) := \sum_{b \in \mathcal{B}} f_{ub} x_{ub} c_{ub}.$$

*B. MORA criterion*

In line with previous approaches [21]–[23], the underlying assumption behind our criterion is that operators share the cost of deploying and/or maintaining the infrastructure, and the resources received by each operator should be based on the level of its (financial) contribution to the shared network: if an operator contributes twice as much as another, it should roughly get twice the resources. To this end, each operator is assigned a *network share* $s_o \in [0, 1]$, to represent its level of contribution to the network. Without loss of generality, these shares are normalized so that $\sum_{o \in \mathcal{O}} s_o = 1$.

The proposed criterion allocates resources across operators dynamically, tracking changes in the numbers and locations of operators' mobile users and the associated transmission rates $c_{ub}$. When doing this, we need to make sure that (i) network resources are fairly shared among the various operators according to their share, and (ii) at the same time, the resources allocated to a given operator are fairly shared among the users of that operator. To achieve this, we follow an approach akin to that in [26][2]: we maximize the overall network utility resulting from aggregating operator utilities, where the utility of an operator is in turn the aggregation of its users' utilities.

To aggregate operators' utilities, we define the overall network utility as the weighted sum of operators' utilities,

$$W(\mathbf{x}, \mathbf{f}) = \sum_{o \in \mathcal{O}} s_o U_o(\mathbf{x}, \mathbf{f}),$$

where, by weighting the utility of an operator by its share, we ensure that operators with larger shares are allocated more resources. We further define the *operator utility* as the sum utility of the operator's users, where a user's utility is logarithmic in its rate, i.e.,

$$U_o(\mathbf{x}, \mathbf{f}) = \frac{1}{|\mathcal{U}_o|} \sum_{u \in \mathcal{U}_o} \log(r_u(\mathbf{x}, \mathbf{f})),$$

where the operator utility is normalized by the number of users of the operator to align the utility values of operators with different numbers of users.

By combining the above equations, one can rewrite the network utility as follows:

$$W(\mathbf{x}, \mathbf{f}) = \sum_{o \in \mathcal{O}} \sum_{u \in \mathcal{U}_o} w_u \log(r_u(\mathbf{x}, \mathbf{f})), \quad (1)$$

[2]Reference [26] addresses a similar problem to ours in the context of users and flows, as it aims at allocating resources fairly to users while preserving fairness among the flows of each user.

where the user weights $w_u$ are defined as the operator network share divided by the total number of users of the operator, i.e., $w_u = s_o / |\mathcal{U}_o|$ (in simple terms, the network share of an operator is divided equally amongst its current users).

With the above, we can now formulate the MORA optimization problem as follows:

$$\max_{\mathbf{x}, \mathbf{f}} W(\mathbf{x}, \mathbf{f}), \quad (2a)$$

subject to:

$$r_u(\mathbf{x}, \mathbf{f}) = \sum_{b \in \mathcal{B}} f_{ub} x_{ub} c_{ub}, \quad \forall u \quad (2b)$$

$$\sum_{b \in \mathcal{B}} x_{ub} = 1 \text{ and } x_{ub} \in \{0, 1\}, \quad \forall b, u \quad (2c)$$

$$\sum_{u \in \mathcal{U}} f_{ub} x_{ub} \leq 1 \text{ and } f_{ub} \geq 0, \quad \forall b, u. \quad (2d)$$

In the sequel we shall let $\mathbf{x}^{MORA}, \mathbf{f}^{MORA}$ denote a (possibly not unique) optimal solution to this optimization problem.

At any given time the above optimization corresponds to the weighted *proportional fair* criterion (see e.g. [4]) extended to a multi-operator setting. In such multi-operator setting, individual user utilities within the network utility are not as important as operators' utilities, given by the sum of the respective user utilities. Also, as users' weights depend on operator shares and their associated number of active users, these affect the interaction among the operators' spatial loads when sharing the network.

*C. Properties of MORA Resource Allocation*

Next, we show that the MORA criterion satisfies some desirable properties both in the way base stations' resources are allocated to associated users, and the way users are associated with base stations.

*1) Per-base station resource allocation:* Let us first consider a general setting, where user associations to base stations are *fixed*, to see how MORA allocates base station resources. Let $\mathbf{x}^*$ be the fixed (not necessarily optimal) user to base station association. If we optimize the resource allocation $\mathbf{f}$ for this user association, i.e., $\max_{\mathbf{f}} W(\mathbf{x}^*, \mathbf{f})$ subject to (2b) and (2c), it can be seen from Lemma 5.1 of [9] that the resulting resource allocation is unique and given by $\mathbf{f}^M(\mathbf{x}^*) = (f_{ub}^M(\mathbf{x}^*) : u \in \mathcal{U}, b \in \mathcal{B})$, where

$$f_{ub}^M(\mathbf{x}^*) = \frac{w_u x_{ub}^*}{\sum_{v \in \mathcal{U}} w_v x_{vb}^*}. \quad (3)$$

Further if $\mathbf{x}^* = \mathbf{x}^{MORA}$, then $\mathbf{f}^M(\mathbf{x}^*) = \mathbf{f}^{MORA}$, i.e., we have MORA optimal allocation of network resources.

The above result is fairly intuitive. Users associated with a given base station are allocated resources proportionally to their weights $w_u$. This can be viewed as follows. The share of an operator represents the total budget of the operator. When assigning a weight $w_u = s_o / |\mathcal{U}_o|$ to users, this share is distributed among the operator's users, and hence the user's weight represents the budget of a user. As the resources allocated to a user are inversely proportional to the sum of weights at her base station, the sum of weights can be viewed as the cost of a unit of resource at the base station. Thus, operators with users associated with heavily loaded base stations will have to pay a higher cost (e.g, increase their network share or limit their overall number of users) or receive fewer resources.

The above shows that the number of active users that operators have on the network and their *spatial distribution* will impact the resources the are allocated under MORA. Indeed, allocations across base stations are coupled together through $|\mathcal{U}_o|$, i.e., an operator with a large number of active users will have lower weights and likely lower per-user allocations. At the same time, the resources obtained by an operator heavily depend on the *load* at base stations to which its users will be associated with.

*2) Characteristics of user association:* Next we study the properties of MORA user associations.

The first result, which follows from the optimality of MORA allocations, is that the resulting resource allocations is Pareto-optimal, which means that for any alternative allocation $(\mathbf{x}', \mathbf{f}')$ for which $r_u(\mathbf{x}', \mathbf{f}') > r_u(\mathbf{x}^{MORA}, \mathbf{f}^{MORA})$ for some $u$, we necessarily have $r_v(\mathbf{x}', \mathbf{f}') < r_v(\mathbf{x}^{MORA}, \mathbf{f}^{MORA})$ for some $v \neq u$. Indeed, if this was not the case then $W(\mathbf{x}', \mathbf{f}')$ would be larger than $W(\mathbf{x}^{MORA}, \mathbf{f}^{MORA})$, which contradicts the fact that the optimal MORA allocation $(\mathbf{x}^{MORA}, \mathbf{f}^{MORA})$ maximizes $W(\mathbf{x}, \mathbf{f})$.

Thus, Pareto optimality in this context means that if under some other user association choice, a user sees a higher throughput than that under MORA then there must be another user which sees a lower throughput allocation. Note that this need not always be the case. Consider, for instance, a network with $|\mathcal{U}|$ users, such that the largest $c_{ub}$ of each user corresponds to a different base station. While the optimal allocation would associate each user to the base station with largest $c_{ub}$, a criterion based on local decisions that looks at users one by one may lead to a different association. The above result guarantees that this will not happen under MORA.

While the above shows the optimality of MORA user associations, it leaves the question of whether this is achieved at the detriment of some operator. The following theorem establishes that if an operator could unilaterally change its users' associations to improve its own utility, it would obtain a limited benefit. This suggests that MORA is not harming any operator for the global benefit: if it did so, such an operator would likely obtain a large gain by modifying its users' associations.

**Theorem 1.** *Consider an optimal solution* $(\mathbf{x}^{MORA}, \mathbf{f}^{MORA})$ *to MORA. Suppose operator* $o \in \mathcal{O}$ *unilaterally modifies its users' associations leading to a new overall user association vector* $\mathbf{x}'$, *and that resources at each base station are allocated according MORA's resource allocation* $\mathbf{f}^M(\mathbf{x}')$ *as given by* (3). *Then, the overall change in utility for operator o, denoted by* $\Delta U_o$, *is bounded below by* $-\log(e)$, *i.e.,*

$$\Delta U_o \doteq U_o(\mathbf{x}^{MORA}, \mathbf{f}^{MORA}) - U_o(\mathbf{x}', \mathbf{f}^M(\mathbf{x}')) \geq -\log(e).$$

*Proof.* See the Appendix. □

The above result also shows that if an operator had control over its users' associations, it would not be able to significantly increase its own utility (at the detriment of others) by deviating from MORA optimal user associations.

*D. Gains and Savings*

In the following we evaluate the benefits of having MORA-based infrastructure slicing. To that end, we introduce a simple baseline – *static slicing* (SS), a proxy for not sharing resources at all.

*1) Static Slicing (SS) Baseline:* This baseline assume that each operator contracts for a *fixed* slice/fraction $s_o$ of the network resources at each base station, i.e., the allocation of resources in each station is static and correspond to the assignation of $s_o$ fraction of resources of every station to operator $o$. The operator can of course still optimize its users associations, $\mathbf{x}^o = (x_{ub} : u \in \mathcal{U}_o, b \in \mathcal{B})$, and allocation of its $s_o$ resources among its users, $\mathbf{f}^o = (f_{ub} : u \in \mathcal{U}_o, b \in \mathcal{B})$, so as to maximize its utility. Specifically each operator $o \in \mathcal{O}$ can determine its user association and resource allocations based on:

$$\max_{\mathbf{x}^o, \mathbf{f}^o} \quad U_o(\mathbf{x}^o, \mathbf{f}^o) \quad (4)$$

$$\text{subject to} \quad r_u(\mathbf{x}^o, \mathbf{f}^o) = \sum_{b \in \mathcal{B}} f_{ub} x_{ub} c_{ub}, \ \forall u \in \mathcal{U}_o,$$

$$\sum_{b \in \mathcal{B}} x_{ub} = 1, \ \forall u \in \mathcal{U}_o,$$

$$\sum_{u \in \mathcal{U}_o} f_{ub} x_{ub} \leq s_o, \ \forall b \in \mathcal{B},$$

$$x_{ub} \in \{0,1\}, \ f_{ub} \geq 0, \ \forall b \in \mathcal{B}, \forall u \in \mathcal{U}_o.$$

This is similar to MORA except limited to the operator $o$'s current users $\mathcal{U}_o$ and the resource constraint is restricted only to the fixed slice $s_o$ allocated to the operator at each base station. Although the user associations and resource allocations under static slicing are independently optimized by each operator, we shall let $\mathbf{x}^{SS}, \mathbf{f}^{SS}$ be a (possibly not unique) optimal choice across all operators under static slicing. Also paralleling our discussion of MORA, it is easy to show that if one fixes a feasible user association $\mathbf{x}^*$, (4) is convex and yields resource allocations given by

$$\mathbf{f}^S(\mathbf{x}^*) := (f_{ub}^*(\mathbf{x}^*) \ : \ \forall u \in \mathcal{U}, \forall b \in \mathcal{B}),$$

where

$$f_{ub}^*(\mathbf{x}^*) = \frac{x_{ub}^* s_o}{\sum_{v \in \mathcal{U}_o} x_{vb}^*} \mathbf{1}\{u \in \mathcal{U}_o\}, \quad (5)$$

i.e., this is again a weighted proportionally fair allocation of the operators' slice of the base station resources.

*2) Operator Utility Gains and Protection:* The *overall* network utility under MORA is clearly larger than that under the more constrained allocations possible under SS. This however does not guarantee that a given *operator's* utility under MORA is greater than that under SS. Below we show that for the *same* user association an operator utility under MORA exceeds that under SS, indicating that beyond the overall network utility, we have that each operator is indeed better off. This shows that MORA effectively protects operators when sharing their resources with other operators.

**Theorem 2.** *For a given user association* $\mathbf{x}$*, MORA's resource allocation* $\mathbf{f}^M(\mathbf{x})$ *(see Eq. 3) achieves a higher utility than that of SS given by* $\mathbf{f}^S(\mathbf{x})$ *(see Eq. 5), i.e., for all* $o \in \mathcal{O}$

$$U_o(\mathbf{x}, \mathbf{f}^M(\mathbf{x})) \geq U_o(\mathbf{x}, \mathbf{f}^S(\mathbf{x})).$$

*Proof.* See the Appendix. □

Theorem 2 does not take into account that SS and MORA may choose different user associations. The following result takes this into account giving a bound on the possible operator utility gap.

**Theorem 3.** *The optimal utilities under MORA and SS are such that for all* $o \in \mathcal{O}$

$$U_o(\mathbf{x}^{MORA}, \mathbf{f}^{MORA}) - U_o(\mathbf{x}^{SS}, \mathbf{f}^{SS}) \geq -\log(e).$$

*Proof.* See the Appendix. □

This analysis suggests that there may be cases in which an operator sees a higher utility under SS than MORA, but the additional utility can be no more than $\log(e)$.

*3) Capacity Savings:* Next we consider the capacity savings resulting from operators sharing infrastructure. Specifically we compare the spectrum capacities, i.e., total amount of resource, required to achieve the same *average utility* per operator under MORA and SS. The aim is to give some intuition on the typical savings one might expect and its dependence on the network load, number of operators and their shares. For tractability we will examine a scenario where traffic is spatially homogenous and operators' network shares are proportional to their load.

We consider a network model in which there is a *fixed* total number of users $|\mathcal{U}|$ of which each operator contributes a fixed number of users proportional its network share $s_o$, i.e., $n_o = s_o|\mathcal{U}|$ which are assumed to be integer valued. Each operator's users are randomly (uniformly) distributed amongst the $|\mathcal{B}|$ base stations, so the number of users of operator $o$ associated with base station $b$, is given by a random variable $N_{o,b}$, such that $N_{o,b} \sim \text{Binomial}(n_o, \frac{1}{|\mathcal{B}|})$. The total number of users at base station $b$ is denoted by a random variable $N_b = \sum_{o \in \mathcal{O}} N_{o,b} \sim \text{Binomial}(|\mathcal{U}|, \frac{1}{|\mathcal{B}|})$. We also assume for simplicity that users have the same capacity $c_{ub} = c$ to the base stations with which they associate.

Note that under the above traffic model all users $u$ have the *same* weight $w_u = \frac{s_o}{n_o} = 1/|\mathcal{U}|$. Thus expected overall network utility under MORA is given by:

$$\bar{W} = \mathbb{E}\left[\sum_{o \in \mathcal{O}}\sum_{b \in \mathcal{B}} N_{ob} w_u \log\left(\frac{c}{N_b}\right)\right] = \mathbb{E}\left[\sum_{b \in \mathcal{B}}\sum_{o \in \mathcal{O}} \frac{N_{ob}}{|\mathcal{U}|} \log\left(\frac{c}{N_b}\right)\right]$$

$$= \mathbb{E}\left[\sum_{b \in \mathcal{B}} \frac{N_b}{|\mathcal{U}|} \log\left(\frac{c}{N_b}\right)\right] = \frac{|\mathcal{B}|}{|\mathcal{U}|} \mathbb{E}\left[N_b \log\left(\frac{c}{N_b}\right)\right],$$

where the last equality follows by using the uniformity of traffic across base stations. Moreover, under our model the network utility $\bar{W}$ is the average utility across all users, which by symmetry is equal to the expected utility of a given operator $o$ under MORA, i.e., $\bar{U}_o^{MORA} = \bar{W}$.

Now applying Taylor's approximation to the function $x \log(c/x)$ at $\mathbb{E}[N_b]$ we obtain

$$N_b \log\left(\frac{c}{N_b}\right) \approx \mathbb{E}[N_b]\log\left(\frac{c}{\mathbb{E}[N_b]}\right) + \left[\log\left(\frac{c}{\mathbb{E}[N_b]}\right) - 1\right] \cdot$$

$$(N_b - \mathbb{E}[N_b]) - \frac{1}{2\mathbb{E}[N_b]}(N_b - \mathbb{E}[N_b])^2,$$

which in turn gives

$$\mathbb{E}\left[N_b \log\left(\frac{c}{N_b}\right)\right] \approx \mathbb{E}[N_b]\log\left(\frac{c}{\mathbb{E}[N_b]}\right) - \frac{1}{2\mathbb{E}[N_b]}\text{Var}(N_b).$$

Since $N_b \sim \text{Binomial}(|\mathcal{U}|, \frac{1}{|\mathcal{B}|})$ we have that $\text{Var}(N_b) = \frac{|\mathcal{U}|}{|\mathcal{B}|}(1 - \frac{1}{|\mathcal{B}|}) \approx \frac{|\mathcal{U}|}{|\mathcal{B}|}$, and so

$$\bar{U}_o^{MORA} \approx \log\left(\frac{c}{\mathbb{E}[N_b]}\right) - \frac{|\mathcal{B}|}{2|\mathcal{U}|}. \qquad (6)$$

Let $\Delta_o$ denote the extra capacity that operator $o$ would require under SS to achieve the above utility. The expected utility experienced by operator $o$ under SS is given by

$$\bar{U}_o^{SS} = \mathbb{E}\left[\sum_{b \in \mathcal{B}} \frac{N_{o,b}}{n_o} \log\left(\frac{s_o c(1+\Delta_o)}{N_{o,b}}\right)\right]$$

$$= \frac{|\mathcal{B}|}{n_o}\mathbb{E}\left[N_{o,b}\log\left(\frac{s_o c}{N_{o,b}}\right)\right] + \log(1+\Delta_o).$$

Again using a Taylor expansion this can be approximated as

$$\bar{U}_o^{SS} \approx \log\left(\frac{s_o c}{\mathbb{E}[N_{o,b}]}\right) - \frac{|\mathcal{B}|}{n_o}\frac{\text{Var}(N_{o,b})}{2\mathbb{E}[N_{o,b}]} + \log(1+\Delta_o).$$

Noting that $\text{Var}(N_{o,b}) \approx s_o \frac{|\mathcal{U}|}{|\mathcal{B}|} = \frac{n_o}{|\mathcal{B}|}$ we have that

$$\bar{U}_o^{SS} \approx \log\left(\frac{c}{\mathbb{E}[N_b]}\right) - \frac{|\mathcal{B}|}{2n_o} + \log(1+\Delta_o). \qquad (7)$$

Finally equating the expected utilities, i.e., (6) and (7), we obtain the following estimate of the necessary extra capacity $\Delta_o$ required when static slicing rather than MORA is used:

$$\log(1+\Delta_o) \approx \frac{|\mathcal{B}|}{2n_o} \times (1 - s_o). \qquad (8)$$

where under our traffic load model $n_o = s_o|\mathcal{U}|$.

This result gives a clear intuition on the possible savings resulting from sharing the infrastructure with MORA dynamic slicing. In particular, the savings increase exponentially in the product of two terms. The first is inversely proportional to the average number of users operator $o$ has per base station, i.e., $n_o/|\mathcal{B}|$; indeed, if the operator has a large number of users, its multiplexing gain is already high without sharing the infrastructure, and hence there is little gain from sharing. The second term is large when the operator has a small network share: if its share is high, the operator is using most of the network resources and there is little sharing.

In summary, capacity savings will be highest when infrastructure is shared by a large number of operators each with a small number of users per base station. With current trends towards small cells, the number of users per base station is expected to be small, suggesting that infrastructure sharing may be particularly beneficial.

## III. APPROXIMATION ALGORITHM FOR MORA

The analysis in previous section and simulations to be presented in the sequel suggest that MORA resource allocation across operators not only has desirable characteristics but will make efficient use of resources while protecting operators from one another. Unfortunately, as we show below, the complexity and information overheads associated with doing so for are already high for a static system, and excessive when operators' mobile users and associated channels are subject to constant change. In this section we further discuss the state-of-the-art algorithms to tackle MORA, and then propose an approximation algorithm based on a sequence of theoretical results and insights that support the design.

### A. Complexity and State-of-the-Art Algorithms

The optimization problem underlying MORA is a *non-linear integer programming problem*, which can be shown to be NP-hard and hence there is no polynomial time algorithm unless $P = NP$.

**Theorem 4.** *The MORA problem is NP-hard.*

*Proof.* See the Appendix. □

There have been a number of works in the literature devoted to solving problems similar to MORA. In particular, [9] proposes an approximation algorithm for the single operator case with guaranteed performance bounds. However, their approach is still computationally demanding; indeed, the results in Section IV-D, show that for a network with only 100 users, the algorithm takes 20 seconds on a dual-core 2.8GHz processor. Given that this would need to be executed every time $c_{ub}$ values change or new users enter/leave the network, this seems computationally impractical. Moreover the proposed approach is centralized, so there would be a substantial information overhead to gather the $c_{ub}$ of each user to each potential base stations, given the amount of data and dynamic nature of mobile users.

In the multi-operator setting, [23] proposes an approach based on using a standard non-linear solver to address a problem similar to MORA. Unfortunately the approach is also very complex and centralized. Indeed, our evaluation of this proposal in Section IV-D, shows that the time required to execute this algorithm increases sharply with the number of users, making it impractical at about 50 users. Moreover, [23] does not provide any analytical performance bounds.

In summary, to make dynamic multi-operator resource sharing possible, a new radically simplified approach is required. It should have low computational complexity and be based on distributed operation requiring only local information, to allow near real-time operation.

### B. Algorithm design

In the following, we devise an algorithm for MORA that can be used in practical deployments. In contrast to previous approaches, our algorithm involves a low computational complexity and relies on data that can be gathered from neighboring base stations, allowing for a distributed implementation.[3]

Given the user dynamics, i.e., joining, moving and leaving the network, an offline algorithm that computes an optimal resource allocation for a fixed set of users is impractical. Instead, we will pursue an approach that tracks users dynamics, and occasionally adjusts resource allocations by modifying current or new users' associations. Since reassociations of current users correspond to handoffs, their number should be kept to a minimum. To design such an algorithm, we need to answer the following questions:

- Do we really need to reassociate users?
- Where should users be (re)associated to?
- In which order should users be reassociated?
- How many reassociations do we need?

For each of these questions, in the following we provide some theoretical analysis that eventually leads to our proposed algorithm. In all cases, once a user association $\mathbf{x}$ is set, resources at each base station are allocated according MORA's resource allocation $\mathbf{f}^M(\mathbf{x})$.

*1) Need for reassociations:* Following the standard terminology of online algorithms, we say that an algorithm is *online* if, upon a user joining the network, it only decides how to associate the new user, without triggering any reassociations of existing users. We say the algorithm is *semi-online* if it can further trigger reassociations of a limited number of users. Thus our first question is whether an online algorithm would suffice. The following theorem suggests that the performance of an online algorithm can be arbitrarily bad, motivating us to consider semi-online approaches.

**Theorem 5.** *Consider an online algorithm that triggers no reassociations of existing users. Let $(\mathbf{x}', \mathbf{f}')$ denote the solution resulting from this algorithm and $(\mathbf{x}^{MORA}, \mathbf{f}^{MORA})$ a MORA optimal solution. Then, $W(\mathbf{x}^{MORA}, \mathbf{f}^{MORA}) - W(\mathbf{x}', \mathbf{f}')$ cannot be bounded.*

*Proof.* See the Appendix. □

*2) Criterion for (re)associations:* Next we address the question regarding how to associate, or reassociate, users to base stations. In particular, consider a *Distributed Greedy algorithm* wherein we iteratively examine (in arbitrary order) if there is a user which could change her association to increase her rate, and if this is the case, she chooses to re-associate with the base station providing the largest rate. The following result characterizes the performance of this algorithm *if* an equilibrium is reached.

**Theorem 6.** *Let $(\mathbf{x}', \mathbf{f}')$ be an equilibrium allocation for the Distributed Greedy algorithm, and $(\mathbf{x}^{MORA}, \mathbf{f}^{MORA})$ a MORA optimal solution, then*

$$W(\mathbf{x}', \mathbf{f}') \geq W(\mathbf{x}^{MORA}, \mathbf{f}^{MORA}) - \log(e).$$

*Proof.* See the Appendix. □

---

[3]Note that, while the algorithm implementation is distributed, the logic is centralized: i.e., we assume that the algorithm is run centrally by a single entity, without the intervention of the different operators.

Note we have not established the convergence of this algorithm. The convergence of this type of algorithms has received substantial attention in the literature [13], [27], [28]. Indeed, since the throughput of user $u$ is an increasing function of $c_{ub}/\sum_{v \in \mathcal{U}} w_v x_{vb}$, the Distributed Greedy algorithm can be viewed as a congestion game in which the load at a base station is given by the sum of weights of the users at the base station, $l_b = \sum_{v \in \mathcal{U}} w_v x_{vb}$, and a user seeks to minimize $a_{ub} l_b$, where $a_{ub} = 1/c_{ub}$. This game falls in the category of a singleton weighted congestion game with player-specific multiplicative constants and linear variable cost. Based on the lack of a counter-example and the existence of polynomial-time algorithms for special cases, [27] conjectures that this type of games have an equilibrium (see Conjecture 3.7 of [27]). Based on the simulations we have run for numerous instances of the game, we further conjecture that the Distributed Greedy algorithm (which implements a best response dynamics) converges to this equilibrium.

With the above conjecture, the Distributed Greedy approach represents a valuable theoretical contribution from an algorithmic perspective, as it solves MORA problem with a performance bound similar to state-of-the-art solutions but in a distributed manner and with much lower computational complexity (see the results of Section IV-D).

In particular, Distributed Greedy satisfies $W(\mathbf{x}', \mathbf{f}') \geq W(\mathbf{x}^{MORA}, \mathbf{f}^{MORA}) - \log(e)$, while [9] proposes an algorithm that provides a throughput larger than $r_u(\mathbf{x}^{MORA}, \mathbf{f}^{MORA})/(2+\epsilon)$ to all users, which translates into $W(\mathbf{x}', \mathbf{f}') \geq W(\mathbf{x}^{MORA}, \mathbf{f}^{MORA}) - \log(2+\epsilon)$; hence, the algorithm of [9] provides only a slightly tighter bound than Distributed Greedy.

*3) Order of reassociations:* While our analysis of the Distributed Greedy algorithm suggests a user should (re)associate to maximize her rate, it does not indicate in which order user reassociations should be considered to speed up convergence. To address this, we consider the *Greedy Largest Gain algorithm*, which operates as the Distributed Greedy algorithm but at each iteration updates the association of the user achieving the highest gain, i.e., the one achieving the largest $r_u^{new}/r_u^{old}$, where $r_u^{old}$ is the user's current throughput and $r_u^{new}$ is the throughput she would receive under the improved association.

The following theorem shows that the Greedy Largest Gain algorithm exhibits a desirable convergence property. In particular, one can guarantee that at each iteration the network utility increases until it reaches $W(\mathbf{x}^{MORA}, \mathbf{f}^{MORA}) - 2\log(e)$, and from then on it never decreases below $W(\mathbf{x}^{MORA}, \mathbf{f}^{MORA}) - (2+\max_u w_u)\log(e)$.[4] Note that Distributed Greedy does not exhibit this kind of behavior: if we select users in an arbitrary order, the network utility may decrease at any iteration (as the increase in utility of the reassociated user may be smaller than the decrease experienced by the other users).

**Theorem 7.** *Let $(\mathbf{x}^i, \mathbf{f}^i)$ be the solution at the $i^{th}$ iteration of the Greedy Largest Gain algorithm and $(\mathbf{x}^{MORA}, \mathbf{f}^{MORA})$ a MORA optimal solution. Then $W(\mathbf{x}^i, \mathbf{f}^i)$ increases at each iteration until $W(\mathbf{x}^i, \mathbf{f}^i) \geq W(\mathbf{x}^{MORA}, \mathbf{f}^{MORA}) - 2\log(e)$, and thereafter it never decreases below $W(\mathbf{x}^{MORA}, \mathbf{f}^{MORA}) - (2+\max_u w_u)\log(e)$.*

*Proof.* See the Appendix. □

Note that while this is not exactly a convergence result, it asserts that resource allocations are such that the network utility is close to optimal.

*4) Proposed algorithm: Greedy Local Largest Gain.* Based on the above considerations we now propose our algorithm for MORA, the *Greedy Local Largest Gain* algorithm. We shall first describe how it operates at a high level, and then provide a more detailed algorithmic description. When a user joins the network, she greedily joins the base station providing the largest throughput. However, as we have seen, we may need to consider triggering user reassociations. To limit their number and associated handoffs overheads we constrain these to at most $m$. For the first $m-1$ reassociations, users choose the base station that provides the largest throughput, but in the $m^{th}$ the user chooses the base station so as to maximize the network utility $W(\mathbf{x}, \mathbf{f})$. In each of these steps, we select which user to reassociate (if any) based on Greedy Largest Gain criterion, but instead of considering all users in the network, involving possibly a high overhead, we restrict the selection *locally* to users associated with only two base stations (see below).

At a more detailed level, the algorithm needs to consider the following cases: ($i$) a user joins the network, ($ii$) leaves, or ($iii$) changes her location. The algorithm for a joining user is detailed in the pseudocode of next page. The rationale is as follows. In the optimal allocation, users are somehow balanced among base stations, users' weights playing a role in this balance. When a new user joins the network, the balance is broken and the base station with which the user associates may have too many users. Hence, in the first step we reassociate one of the users of this base station. In the next step, the base station that received the reassociated user may have too many users; however, depending on the weights of the joining and reassociated users, the original base station may still have too many users as well. Hence, we consider the users from the two base stations as candidates for reassociation. We repeat this, considering users from two base stations, in the subsequent steps. Finally, in the last step, to avoid that the reassociation of a user harms the overall performance, we select the base station association that maximizes the overall network utility rather than the throughput of the reassociated user.

When a user leaves the network, the algorithm is quite similar (pseudocode omitted for space reasons). When she moves, her $c_{ub}$ values to the neighboring base stations may change; if, as a result of these changes, at some point the user would receive a larger throughput in a new base station, we reassociate her to this base station. Then, the old base station executes the same algorithm as when a user leaves the network while the new base station executes the algorithm corresponding to a joining user.

*5) Controlling the number of reassociations:* The remaining question is how to set the limit on the number of reassociations $m$, which determines the trade-off between the performance of the algorithm and reassociation overhead. Such trade-offs have been analyzed for a similar setting in [29], which aims to distribute tasks among servers (where each task can only be associated to a restricted set of servers) in such a way that the maximum load across all servers is minimized. This problem is similar to ours, with tasks and servers corresponding to users and base stations respectively, in the particular case where all users have the same $w_u$ and $c_{ub}$. Not unlike their setting, the performance in this case is optimized when base station loads are as balanced as possible (i.e., the highest load is minimized). According to the analysis of [29], the performance in terms of the highest load with our algorithm (which has a limit of $m$ reassociations) over the highest load with the optimal algorithm (with no constraint $m$) is given by $O(e^{1-\frac{m}{\ln|\mathcal{B}|}})$. This shows that algorithm's performance improves rapidly (exponentially) in $m$, and suggests a small $m$ suffices to achieve near-optimal network utility.

---

[4]Note that, if all operators have a large number of users, we have that $2 + \max_u w_u \approx 2$.

**Algorithm 1:** GLLG user joining.
**Definitions**:
$r_{v,b}$ : throughput of user $v$ if she associates to $b$;
$r_v$ : current throughput of user $v$;
$\mathcal{U}_b$ : set of users associated to $b$, $(u \in \mathcal{U}$ s.t. $x_{u,b} = 1)$;
$\mathcal{U}_{\{c \cup p\}}$ : set of users associated to $c$ or $p$;
$W_{u,q}$ : network utility if user $u$ associates to $q$;
**Input**: x
**User $v$ joins the network**:
$b' = \arg\max_{b \in \mathcal{B}} r_{v,b}$;
$x_{v,b'} = 1 \leftarrow$ Associate user $v$ with base station $b'$;
$[u^*, p^*] = \arg\max_{(u,p) \in \mathcal{U}_{b'} \times \mathcal{B}} \frac{r_{u,p}}{r_u}$;
**if** $r_{u^*,p^*}/r_u > 1$ **then**
  Associate user $u^*$ with base station $p^*$, $x_{u^*p^*} = 1$;
**else**
  stop
$c = p^*$ (current base station);
$p = b'$ (previous base station);
**for** $m - 1$ *times* **do**
  $[u^*, q^*] = \arg\max_{(u,q) \in \mathcal{U}_{\{c \cup p\}} \times \mathcal{B}} \frac{r_{u,q}}{r_u}$;
  **if** $r_{u^*,q^*}/r_u > 1$ **then**
    Associate user $u^*$ with base station $q^*$, $x_{u^*q^*} = 1$;
    $c \leftarrow q^*$; $p \leftarrow$ previous base station of user $u^*$;
  **else**
    stop
$W \leftarrow$ current network utility;
$[u^*, q^*] = \arg\max_{(u,q) \in \mathcal{U}_{\{c \cup p\}} \times \mathcal{B}} \frac{W_{u,q}}{W}$;
**if** $W_{u^*,q^*}/W > 1$ **then**
  Associate user $u^*$ with base station $q^*$, $x_{u^*q^*} = 1$;

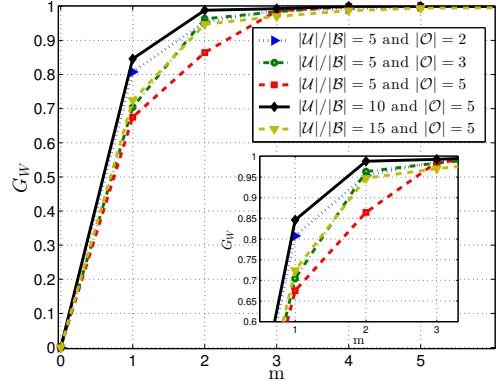

Fig. 1. Normalized utility gain as a function of $m$.

To further explore the impact of $m$ on network utility, we present the following simulation results (see Section IV for a description of the simulation setup). Here, $W(m)$ is the network utility achieved for a given $m$ value, $W(\infty)$ is the utility with unconstrained overhead, $W(0)$ is the utility with no reassociations, and $G_W(m) \doteq 1 - \frac{W(m)-W(\infty)}{W(0)-W(\infty)}$ represents the normalized utility gain with $m$ reassociations, showing how close we get to the unconstrained overhead utility. Fig. 1 depicts this gain as a function of $m$ for different scenarios. As can be seen, utility gains increase very sharply. Furthermore, for $m = 3$ the gains are already very close to their maximum value; based on this, we set $m$ equal to 3 (this is indeed the value used in the experiments of Section IV).

## IV. PERFORMANCE EVALUATION

Next, we evaluate the performance of our proposed approach. The mobile network scenario considered is based on the IMT Advanced evaluation guidelines for dense 'small cell' deployments [30]. It consists of base stations with an intersite distance of 200 meters in a hexagonal cell layout with 3 sector antennas (thus in this setting users will associate with sectors rather than the base stations we used in our algorithm description). The Signal Interference to Noise Ratio (SINR) is computed as in [25], $\text{SINR}_{ub} = P_b g_{ub}/(\sum_{k \in \mathcal{B}, k \neq b} P_k g_{uk} + \sigma^2)$, where $P_b$ is the transmit power and $g_{ub}$ denotes the channel gain between user $u$ and base station $b$, which includes path loss, shadowing, fast fading and antenna gain. Following [30], we set $P_b = 41$ dBm, $\sigma^2 = -104$dB, path loss equal to $36.7 \log_{10}(\text{dist}) + 22.7 + 26 \log_{10}(f_c)$ for carrier frequency $f_c = 2.5$GHz, and antenna gain of 17 dBi. The shadowing factor is given by a log-normal function with a standard deviation of 8dB (as in [25]) updated every second, and fast fading follows a Rayleigh distribution dependent of the user speed and the angle of incidence (as in [31]). Achievable rates are then computed with the Shannon formula, $\text{BW} \log_2(1 + \overline{\text{SINR}}_{ub})$, for the the average $\overline{\text{SINR}}_{ub}$ given by pathloss and shadowing [24] and a channel bandwidth of $\text{BW} = 10\text{MHz}$ [24]. Finally, the modulation-coding scheme is selected according to the $\overline{\text{SINR}}_{ub}$ thresholds reported in [32]. Unless otherwise stated, users move according to the Random Waypoint Model (RWP), network size $|\mathcal{B}|$ is 57 sectors, all operators have the same share, the number of users of the operator $o$ is proportional to $s_o$, i.e., $|\mathcal{U}_o| = |\mathcal{U}| \cdot s_o$, and confidence intervals are below 1%.

### A. Utility gains

We start by evaluating the gains in terms of the overall network utility. We consider a scenario with a user density of

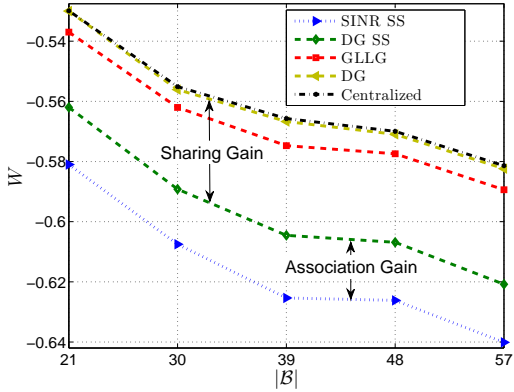

Fig. 2. Utility gains for different approaches as a function of the network size.

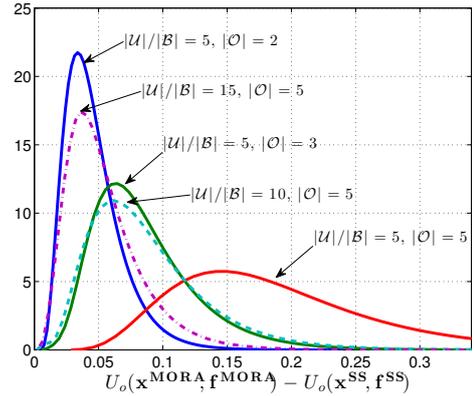

Fig. 3. Distribution for operator utility gain of MORA vs SS.

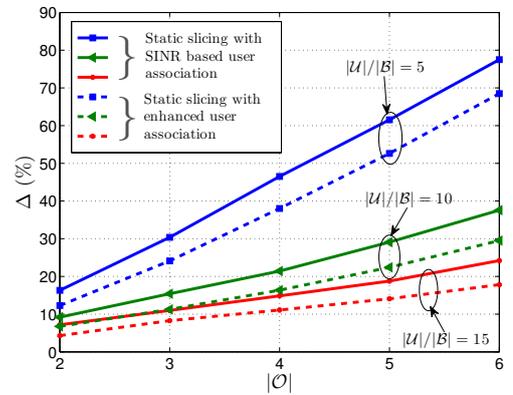

Fig. 4. Capacity savings for different scenarios as a function of the number of operators.

10 users/sector and 3 operators, and plot $W(\mathbf{x}, \mathbf{f})$ as a function of the network size $|\mathcal{B}|$. In this setting, we compare the performance of our algorithm for dynamic sharing, *Greedy Local Largest Gain* ('GLLG'), against the following approaches:

i) *SINR-based Static Slicing* ('SINR SS'): the resources of each sector are statically divided among operators and users associate with the based station with highest SINR;
ii) *Distributed Greedy Static Slicing* ('DG SS'): resources are also sliced statically and user associations follow the Distributed Greedy algorithm discussed in Section III-B2;
iii) *Distributed Greedy* ('DG'): this is the algorithm for dynamic sharing presented in Section III-B2;
iv) *Centralized* ('Centralized'): this is the centralized algorithm proposed in [9].

The results are exhibited in Fig. 2. We draw the following conclusions: (i) significant gains result from both improving user association (DG SS vs. SINR SS) and sharing resources dynamically (DG vs. DG SS); (ii) the Distributed Greedy approach of Section III-B2 performs almost at the same level of the baseline approach of [9] (DG vs. Centralized); and (iii) the proposed approach performs closely to these two approaches, although it pays a small price for reducing the handoff overheads (GLLG vs. DG). Note that the purpose of this results is to benchmark different approaches and the descend trend exhibited in Fig. 2 is caused only by the evaluation in different sized networks.

In addition to the overall network gain, it is also interesting to look at the gains of the individual operators. Theorem 3 showed that the difference in an operator's utility under MORA and SS exceeds $-\log(e)$, but in fact we expect it to be positive. Fig. 3 shows the probability density function for the sampled difference of operators utility under MORA and SS over multiple simulation runs, for different numbers of operators $|\mathcal{O}|$ and user densities $|\mathcal{U}|/|\mathcal{B}|$. As can be seen it is indeed positive, which confirms that MORA effectively protects all operators ensuring gains to all of them. As expected, such gains are higher for scenarios with more operators and lower loads.

### B. Capacity savings

We next evaluate the benefits of our approach to operators based on the capacity savings they would achieve. Specifically, consider a network operated under our algorithm for dynamic sharing, where the capacity (i.e., total amount of resource) of each base station is given by $C_{GLLG}$, and let $C_{baseline}$ be the base stations' capacity required to achieve the same network utility under two baselines: (a) static slicing with SINR-based user association, and (b) static slicing with enhanced user association (i.e., using our algorithm for user association). These two baselines allow us to study the potential gains earned due to a smarter user association and the gains achieved by dynamic resource sharing. Fig. 4 illustrates the corresponding capacity savings, computed as $\Delta = (C_{baseline} - C_{GLLG})/C_{GLLG}$, for different numbers of operators, $|\mathcal{O}| \in \{2, \ldots, 6\}$, and three different user densities, $|\mathcal{U}|/|\mathcal{B}| = 5$ (low density), $|\mathcal{U}|/|\mathcal{B}| = 10$ (medium) and $|\mathcal{U}|/|\mathcal{B}| = 15$ (high). The results show that substantial gains can be realized.

The above results show that gains increase with the number of operators and decrease with per-sector user load. This can be expected intuitively since when the user distribution is denser, the amount of users per station, tend to equalize. By the contrary, with light loads, MORA dynamic resource allocation

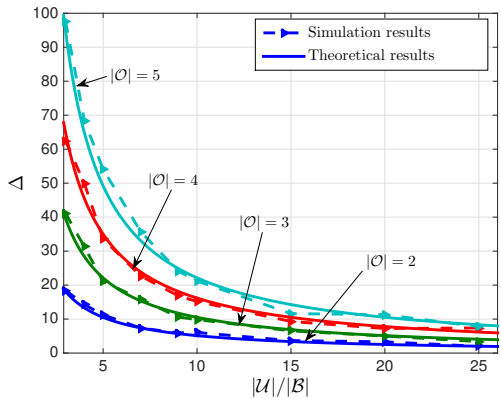

Fig. 5. Validation of the theoretical results on capacity savings.

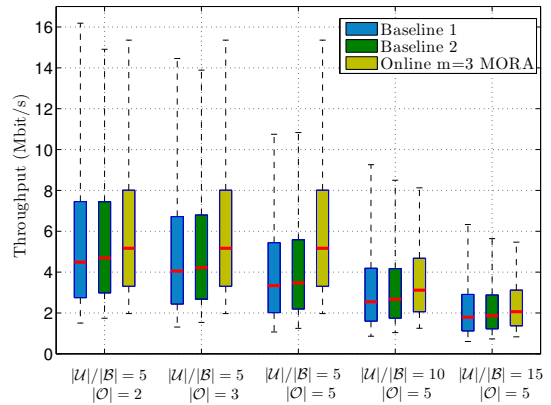

Fig. 6. Improvement on the user throughput.

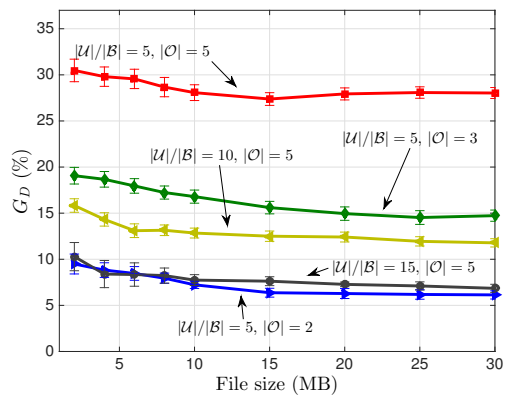

Fig. 7. Improvement on the file download time for different file sizes.

exploit statistical multiplexing in the network, leading to higher gains.

As a consequence of regarding the results from a different angle, results show that in an scenario with operators with different shares, the operators with smaller shares are more benefited from MORA, as suggested in Section II-D3. In order to gain additional insight into the impact of these factors, Fig. 5 displays the influence of the share of the operator ($s_o$) and the average load per base station sector $|\mathcal{U}|/|\mathcal{B}|$ in the percent of extra capacity required to achieve the same utility ($\Delta$) with the static slicing with enhanced user association baseline. Results are also compared with the analytical result of Section II-D3, confirming that the theoretical analysis result holds in real conditions.

### C. User performance

To illustrate the gains from a user perspective, we compare the per-user throughput achieved by our approach against the two baselines: static slicing with SINR-based user association ('Baseline 1'), and static slicing with enhanced user association ('Baseline 2'). The resulting box-and-whisker plots are shown in Fig. 6 for different user densities and numbers of operators. We observe that our approach provides substantial gains both in terms of the median values as well as the various percentiles. Furthermore, similarly to the above, gains increase with the number of operators but decrease with per-sector user load. To complement the previous results, we compare the file download times achieved by our approach against a baseline scenario (static slicing with enhanced user association), when base stations have the same capacity in both cases and users are constantly downloading files. Let us define the file download time gain as $G_D = (D_{SS} - D_{GLLG})/D_{SS}$, where $D_{SS}$ is the average file download time with the static slicing approach and $D_{GLLG}$ with ours. The gains achieved are shown in Fig. 7 as a function of the file download size, for different user densities and numbers of operators. We observe the gains are substantial, and fairly independent of the file size.

### D. Computational complexity

As mentioned in Section III-A, one of the key advantages of the proposed approach over the state-of-the-art is its reduced computational complexity. To quantify this, we have measured the time required to execute the following algorithms in a dual-core 2.8GHz processor: ($i$) our algorithm for dynamic sharing ('GLLG'); ($ii$) the Distributed Greedy approach of Section III-B2, which has unconstrained overhead ('DG'); ($iii$) the centralized algorithm of [9] ('Centralized'); and ($iv$) the non-linear solver used by [23] ('Non-linear Solver'). Fig. 8 shows the resulting execution times (in seconds) as a function of the number of users for a fixed network size $|\mathcal{B}| = 57$ and $|\mathcal{O}| = 4$ operators. The results confirm that the algorithms of [23] and [9] are impractical, especially if we take into account that they have to be triggered every time the channel quality of a user changes. By contrast, the execution time of our Distributed Greedy algorithm remains very low, and it remains even lower for our GLLG approach (due to the constraint that GLLG imposes on the number of handovers).

### E. Impact of non-uniform load distributions

All the results shown so far have been based on the RWP mobility model, which is known to distribute load uniformly across space. To understand the impact of non-uniform load distributions, we have evaluated the capacity savings over a baseline (static slicing with enhanced user association) under the SLAW model [33], which is a non-uniform human walk mobility model. To show different levels of non-uniformity, we have parameterized the SLAW model

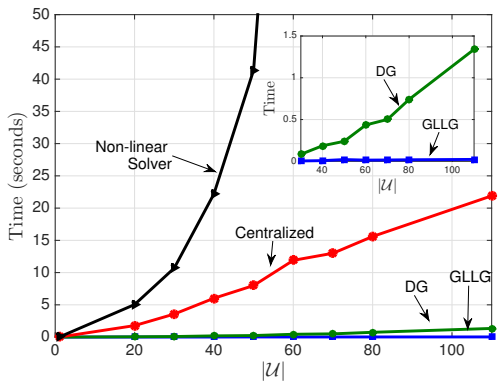

Fig. 8. Computational complexity of our approaches and state-of-the-art algorithms.

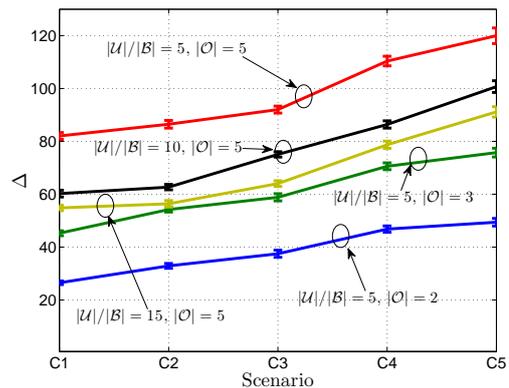

Fig. 10. Capacity savings for different levels of non-uniformity when operators follow different patterns.

## V. CONCLUSIONS

In this paper we have addressed the problem of multi-tenant RAN resource slicing. While there has been substantial work towards addressing this problem, most has focused on architectural issues, leaving algorithmic aspects open to consideration. The design of algorithms for dynamic resource sharing across slices is challenging as it involves user association decisions (which is a difficult problem in itself) as well as multi-operator sharing policies. Our main contribution has been to show that, despite its complexity, it is possible to design practical solutions that scale to large networks and can track network load dynamics. Indeed, our analytical results provide strong evidence that the resulting allocations are near-optimal, and our simulations confirm robust benefits to operators (in terms of capacity savings) as well as to users (in terms of improved performance).

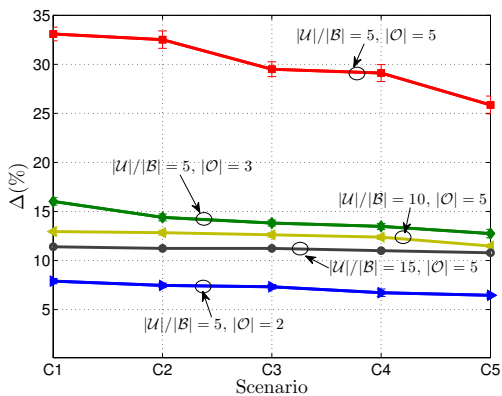

Fig. 9. Capacity savings for different levels of non-uniformity under the SLAW mobility model.

with five different configurations of increasing non-uniformity, from $C1$ to $C5$, whose parameters {waypoints, clustering range, alpha distance, inverse self-similarity} have been set as follows: $C1 = \{100, 20, 5, 0.95\}$, $C2 = \{85, 40, 4.5, 0.85\}$, $C3 = \{75, 60, 4, 0.75\}$, $C4 = \{65, 80, 3.5, 0.65\}$ and $C5 = \{50, 100, 3, 0.55\}$. The results, given in Fig. 9, show that (as expected) capacity savings decrease if loads are non-uniform since multiplexing gains are reduced when users concentrate around some areas. However, the decrease is very gradual, which shows that non-uniformity has a limited impact on the resulting gains.

It is worth noting that the above experiment assumed that all operators follow the same mobility pattern, by using the same instance of the SLAW model to generate the users hotspots; if different patterns are assumed for different operators (which may be the case for instance if we consider services of different nature) by using different instances of the SLAW model to generate the users hotspots, then gains increase (rather than decrease) with non-uniformity, as each operator may have its users concentrated in different hotspots, thereby maximizing the benefit from resource sharing. This is shown by the results given in Fig. 10.

## APPENDIX

*Proof of Theorem 1:* Let $(\mathbf{x}^*, \mathbf{f}^*) = (\mathbf{x}^{MORA}, \mathbf{f}^{MORA})$ be a (possibly not unique) optimal solution to MORA. Since $(\mathbf{x}^*, \mathbf{f}^*)$ is optimal, we have :

$$W(\mathbf{x}^*, \mathbf{f}^*) - W(\mathbf{x}', \mathbf{f}^M(\mathbf{x}')) = \\ \sum_{p \in \mathcal{O}} U_p(\mathbf{x}^*, \mathbf{f}^*) - U_p(\mathbf{x}', \mathbf{f}^M(\mathbf{x}')) \geq 0$$

Thus,

$$\sum_{p \in \mathcal{O} \setminus \{o\}} U_p(\mathbf{x}^*, \mathbf{f}^*) - U_p(\mathbf{x}', \mathbf{f}^M(\mathbf{x}')) \geq$$

$$U_o(\mathbf{x}', \mathbf{f}^M(\mathbf{x}')) - U_o(\mathbf{x}^*, \mathbf{f}^*) = -\Delta U_o.$$

Now let $\mathcal{U}_{o,b}^*$ denote the set of users of operator $o$ connected to base station $b$ under user association $\mathbf{x}^*$, and $\mathcal{U}_{o,b}'$ the same for user association $\mathbf{x}'$. Furthermore, let $\omega_u = 1/|\mathcal{U}_o|$ for $u \in \mathcal{U}_o$, $\omega_{ob}^* = |\mathcal{U}_{o,b}^*|/|\mathcal{U}_o|$ and $\omega_{ob}' = |\mathcal{U}_{o,b}'|/|\mathcal{U}_o|$. Note that for $p \neq o$, $\mathcal{U}_{p,b}^* = \mathcal{U}_{p,b}'$ and $\omega_{p,b}^* = \omega_{p,b}'$. Then,

$$\sum_{p \in \mathcal{O} \setminus \{o\}} U_p(\mathbf{x}^*, \mathbf{f}^*) - U_p(\mathbf{x}', \mathbf{f}^M(\mathbf{x}')) =$$

$$= \sum_{p \in \mathcal{O} \setminus \{o\}} \sum_{b \in \mathcal{B}} \Bigg( \sum_{u \in \mathcal{U}_{p,b}^*} \omega_u \log\Big(\frac{c_{ub}\omega_u}{\sum_{q \in \mathcal{O}} \omega_{qb}^*}\Big) - \\ \sum_{u \in \mathcal{U}_{p,b}'} \omega_u \log\Big(\frac{c_{ub}\omega_u}{\sum_{q \in \mathcal{O}} \omega_{qb}'}\Big) \Bigg)$$

$$= \sum_{b \in \mathcal{B}} \sum_{p \in \mathcal{O} \setminus \{o\}} \sum_{u \in \mathcal{U}_{p,b}^*} \omega_u \log\Big(\frac{c_{ub}\omega_u}{\sum_{q \in \mathcal{O}} \omega_{qb}^*}\Big) - \quad (9)$$

$$\omega_u \log\Big(\frac{c_{ub}\omega_u}{\sum_{q \in \mathcal{O}} \omega_{qb}'}\Big)$$

$$= \sum_{b \in \mathcal{B}} \Big( \sum_{p \in \mathcal{O} \setminus \{o\}} \sum_{u \in \mathcal{U}_{p,b}^*} \omega_u \Big) \log\Big(\frac{\sum_{q \in \mathcal{O}} \omega_{qb}'}{\sum_{q \in \mathcal{O}} \omega_{qb}^*}\Big) \quad (10)$$

$$= \sum_{b \in \mathcal{B}} \Big( \sum_{p \in \mathcal{O} \setminus \{o\}} \omega_{p,b}^* \Big) \log\Big(\frac{\sum_{q \in \mathcal{O} \setminus \{o\}} \omega_{qb}' + \omega_{ob}'}{\sum_{q \in \mathcal{O} \setminus \{o\}} \omega_{qb}^* + \omega_{ob}^*}\Big)$$

$$\leq \sum_{b \in \mathcal{B}} \Big( \sum_{p \in \mathcal{O} \setminus \{o\}} \omega_{p,b}^* \Big) \log\Big(\frac{\sum_{q \in \mathcal{O} \setminus \{o\}} \omega_{qb}' + \omega_{ob}'}{\sum_{q \in \mathcal{O} \setminus \{o\}} \omega_{qb}^*}\Big)$$

$$= \sum_{b \in \mathcal{B}} \omega_{ob}' \log\left(\frac{\sum_{q \in \mathcal{O} \setminus \{o\}} \omega_{qb}^*}{\omega_{ob}'} + 1\right)^{\frac{\sum_{q \in \mathcal{O} \setminus \{o\}} \omega_{qb}^*}{\omega_{ob}'}}$$

$$\leq \sum_{b \in \mathcal{B}} \omega_{ob}' \log e = \log e \quad (11)$$

Note that (9) holds since $\mathcal{U}_{p,b}^* = \mathcal{U}_{p,b}'$ for $p \neq o$, (10) is obtained by using properties of the logarithm, and inequality (11) holds since $((x+1)/x)^x < e$ for $x \geq 0$. □

*Proof of Theorem 2:* For a given user association $\mathbf{x}$ the utility of operator $o$ under SS is maximized when the resource blocks of each operator at each base station are equally distributed among the operator's users. This yields

$$U_o(\mathbf{x}, \mathbf{f^S}(\mathbf{x})) =$$
$$= \sum_{b \in \mathcal{B}} \sum_{u \in \mathcal{U}_o} \frac{1}{|\mathcal{U}_o|} x_{ub} \log\left(\frac{1}{\sum_{b \in \mathcal{B}} \sum_{v \in \mathcal{U}_o} x_{vb}} \frac{s_o}{\sum_{o' \in \mathcal{O}} s_{o'}} c_{ub}\right),$$
$$= \frac{1}{s_o} \sum_{b \in \mathcal{B}} \sum_{u \in \mathcal{U}_o} w_u x_{ub} \log\left(\frac{1}{\sum_{b \in \mathcal{B}} \sum_{v \in \mathcal{U}_o} x_{vb}} \frac{s_o}{\sum_{o' \in \mathcal{O}} s_{o'}} c_{ub}\right),$$

where the weights are $w_u = \frac{s_o}{|\mathcal{U}_o|}$, $u \in \mathcal{U}_o$.

If we multiply the numerator and denominator inside the log() by $w_u$, and take into account that $w_u = w_v$ for $u, v \in \mathcal{U}_o$ and $\sum_{o' \in \mathcal{O}} s_{o'} = 1$, the above can be rewritten as

$$U_o(\mathbf{x}, \mathbf{f}^S(\mathbf{x})) = \frac{1}{s_o} \sum_{b \in \mathcal{B}} \sum_{u \in \mathcal{U}_o} w_u x_{ub} \log(w_u c_{ub}) -$$

$$\frac{1}{s_o}\sum_{b\in\mathcal{B}}\sum_{u\in\mathcal{U}_o} w_u x_{ub}\log\left(\frac{\sum_{b\in\mathcal{B}}\sum_{v\in\mathcal{U}_o} w_v x_{vb}}{s_o}\right).$$

The utility of operator $o$ with MORA allocation is given by

$$U_o(\mathbf{x},\mathbf{f}^M(\mathbf{x}))=\frac{1}{s_o}\sum_{b\in\mathcal{B}}\sum_{u\in\mathcal{U}_o} w_u x_{ub}\log\left(\frac{w_u c_{ub}}{\sum_{v\in\mathcal{U}} w_v x_{vb}}\right),$$

which can be rewritten as

$$U_o(\mathbf{x},\mathbf{f}^M(\mathbf{x}))=\frac{1}{s_o}\sum_{b\in\mathcal{B}}\sum_{u\in\mathcal{U}_o} w_u x_{ub}\log(w_u c_{ub})-$$
$$\frac{1}{s_o}\sum_{b\in\mathcal{B}}\sum_{u\in\mathcal{U}_o} w_u x_{ub}\log\left(\sum_{v\in\mathcal{U}} w_v x_{vb}\right).$$

From the above, if we can show that

$$\sum_{b\in\mathcal{B}}\sum_{u\in\mathcal{U}_o} w_u x_{ub}\log\left(\frac{\sum_{b\in\mathcal{B}}\sum_{v\in\mathcal{U}_o} w_v x_{vb}}{s_o}\right)\geq$$
$$\sum_{b\in\mathcal{B}}\sum_{u\in\mathcal{U}_o} w_u x_{ub}\log\left(\sum_{v\in\mathcal{U}} w_v x_{vb}\right), \quad (12)$$

the theorem is proved.

To show the above, we consider the maximization of function $\sum_{b\in\mathcal{B}} y_b\log(x_b)$ over $x_b$ subject to $\sum_{b\in\mathcal{B}} x_b = 1$. By applying Lagrange multipliers, it can be easily seen that this function is maximized for $x_b = y_b/\sum_{b'\in\mathcal{B}} y_{b'}$. Since both the left and right-hand sides of (12) conform to this constrained optimization problem, and the left-hand side of (12) corresponds to its optimal solution, the inequality of (12) follows. □

*Proof of Theorem 3:* Let $(\mathbf{x}^{MORA},\mathbf{f}^{MORA})$ and $(\mathbf{x}^S,\mathbf{f}^{SS})$ denote optimal allocations under MORA and SS. Suppose we choose $\mathbf{x}'$ as a user association where $x'_u = x_u^{SS}$ for $u\in\mathcal{U}_o$ and $x'_u = x_u^{MORA}$ otherwise. Then by Theorem 1 we have that

$$U_o(\mathbf{x}^{MORA},\mathbf{f}^{MORA})-U_o(\mathbf{x}',\mathbf{f}^M(x'))\geq -\log(e).$$

Furthermore, from Theorem 2 we have

$$U_o(\mathbf{x}',\mathbf{f}^M(\mathbf{x}'))\geq U_o(\mathbf{x}',\mathbf{f}^S(\mathbf{x}')).$$

Combining the above two equations we get

$$U_o(\mathbf{x}^{MORA},\mathbf{f}^{MORA})-U_o(\mathbf{x}',\mathbf{f}^S(\mathbf{x}'))\geq -\log(e),$$

Note that under SS the utility of an operator is independent of the user associations of the other operators, thus given our choice of $\mathbf{x}'$ we have that

$$U_o(\mathbf{x}',\mathbf{f}^S(\mathbf{x}'))=U_o(\mathbf{x}^{SS},\mathbf{f}^{SS})$$

which combined with the above gives

$$U_o(\mathbf{x}^{MORA},\mathbf{f}^{MORA})-U_o(\mathbf{x}^{SS},\mathbf{f}^{SS})\geq -\log(e)$$

which proves the theorem. □

*Proof of Theorem 4:* The reduction is via the 3-dimensional matching problem which is known to be NP-complete. Recall that the 3-dimensional matching problem is stated as follows. Let us consider disjoint sets $C=\{c_1,\ldots,c_n\}$, $D=\{d_1,\ldots,d_n\}$ and $E=\{e_1,\ldots,e_n\}$, and a family $T=\{T_1,\ldots,T_m\}$ of triples with $|T_i\cap C|=|T_i\cap D|=|T_i\cap E|=1$ for $i=1,\ldots,m$, with $m\geq n$. The question is whether $T$ contains a matching, i.e., a subfamily $T'$ for which $|T'|=n$ and $\cup_{T_i\in T'} T_i = C\cup D\cup E$.

Our reduction is along the lines of [?]. We call the triples that contain $c_j$ *triples of type j*. Let $t_j$ be the number of triples of type $j$ for $j=1,\ldots,n$. Base station $i$ corresponds to the triples $T_i$ for $i=1,\ldots,m$. We create two types of users, element users and dummy users. We have $2n$ element users, $u\in\{1,\ldots,2n\}$, corresponding to the $2n$ elements of $D\cup E$. There are $t_j-1$ dummy users of type $j$ for $j=1,\ldots,n$. Note that the total number of dummy users is $m-n$, $u\in\{2n+1,\ldots,m+n\}$. Element users can connect to the base stations that correspond to a triple that contains this element, with a transmission rate of $R$. Dummy users of type $j$ can connect (also with a transmission of $R$) to the base stations that correspond to triples of type $j$. Element users have a weight $w_u = 1/(2m)$ and dummy users have a weight $w_u = 1/m$.[5] We claim that a matching exists if and only if the network utility with the MORA criterion is $W=(n/m)\log(R/2)+((m-n)/m)\log(R)$. □

*Proof of Theorem 5:* We prove the theorem by means of the following example. Let us consider a scenario with $|\mathcal{B}|$ base stations in which $|\mathcal{B}|^2$ users join the network. All users have the same weight and can associate with any of the $|\mathcal{B}|$ base stations with $c_{ub}=1$. Independently of the criterion followed to associate new users, after all users have joined there must be a base station with at least $|\mathcal{B}|$ users. Now, suppose all users but these $|\mathcal{B}|$ leave the network. For this scenario, the network utility provided by the online algorithm is $W(\mathbf{x}',\mathbf{f}')=\sum_{i=1}^{|\mathcal{U}|}\frac{1}{|\mathcal{B}|}\log(\frac{1}{|\mathcal{B}|}) = -\log(|\mathcal{B}|)$. The optimal solution is that each user associates with a different base station, which yields $W(\mathbf{x}^{MORA},\mathbf{f}^{MORA})=\log(1)$. Thus, we have $W(\mathbf{x}^{MORA},\mathbf{f}^{MORA})-W(\mathbf{x}',\mathbf{f}')=\log(1)+\log(|\mathcal{B}|)$, which grows to $\infty$ as $|\mathcal{B}|\to\infty$.

*Proof of Theorem 6:* Since in an equilibrium of the Distributed Greedy algorithm, each user is associated with the base station that maximizes $r_u$, the following holds for all $u$:

$$\sum_{b\in\mathcal{B}} x'_{ub} w_u \log\left(\frac{w_u c_{ub}}{\sum_{v\in\mathcal{U}} x'_{vb} w_v}\right)\geq$$
$$\sum_{b\in\mathcal{B}} x^*_{ub} w_u \log\left(\frac{w_u c_{ub}}{\sum_{v\in\mathcal{U}} x'_{vb} w_v + w_u}\right) \quad (13)$$

where the base station for which $x'_{ub}=1$ is the one with which user $u$ is associated under Distributed Greedy, and the base station for which $x^*_{ub}=1$ is the one with which it is associated under the optimal allocation (i.e., $\mathbf{x}^*=\mathbf{x}^{MORA}$).

---
[5]These weights can be achieved by considering a scenario with two operators, one with twice as many users as the other, in which those users that are neither element nor dummy users can only connect to one base station that is none of the BS $i$ for $i=1,\ldots,m$.

At the base station for which $x_{ub}^* = 1$ we have $\sum_{v \in \mathcal{U}} x_{vb}^* w_v \geq w_u$, so the following also holds:

$$\sum_{b \in \mathcal{B}} x'_{ub} w_u \log\left(\frac{w_u c_{ub}}{\sum_{v \in \mathcal{U}} x'_{vb} w_v}\right) \geq$$
$$\sum_{b \in \mathcal{B}} x_{ub}^* w_u \log\left(\frac{w_u c_{ub}}{\sum_{v \in \mathcal{U}} x'_{vb} w_v + \sum_{v \in \mathcal{U}} x_{vb}^* w_v}\right).$$

Let us define the load at a base station as the sum of weights of the users at the base station, $l_b = \sum_{v \in \mathcal{U}} w_v x_{vb}$. Then, the above can be rewritten as

$$\sum_{b \in \mathcal{B}} x'_{ub} w_u \log\left(\frac{w_u c_{ub}}{l'_b}\right) \geq \sum_{b \in \mathcal{B}} x_{ub}^* w_u \log\left(\frac{w_u c_{ub}}{l'_b + l_b^*}\right)$$

where $l'_b$ and $l_b^*$ are the load at base station $b$ with the Distributed Greedy algorithm and the optimal allocation, respectively.

From the above it follows that

$$w_u \log(r_u(\mathbf{x}^*, \mathbf{f}^*)) - w_u \log(r_u(\mathbf{x}', \mathbf{f}')) \leq$$
$$\sum_{b \in \mathcal{B}} x_{ub}^* w_u \log\left(\frac{w_u c_{ua}}{l_b^*}\right) - \sum_{b \in \mathcal{B}} x_{ub}^* w_u \log\left(\frac{w_u c_{ua}}{l'_b + l_b^*}\right)$$

where $\mathbf{f}^* = f^M(\mathbf{f}^*)$. The above can be expressed as

$$w_u \log(r_u(\mathbf{x}^*, \mathbf{f}^*)) - w_u \log(r_u(\mathbf{x}', \mathbf{f}')) \leq$$
$$- \sum_{b \in \mathcal{B}} x_{ub}^* w_u \log\left(\frac{l_b^*}{l'_b + l_b^*}\right)$$

Summing the above over all users yields

$$W(\mathbf{x}^*, \mathbf{f}^*) - W(\mathbf{x}', \mathbf{f}') \leq - \sum_{u \in \mathcal{U}} \sum_{b \in \mathcal{B}} x_{ub}^* w_u \log\left(\frac{l_b^*}{l'_b + l_b^*}\right)$$

From the above,

$$W(\mathbf{x}^*, \mathbf{f}^*) - W(\mathbf{x}', \mathbf{f}') \leq - \sum_{b \in \mathcal{B}} \log\left(\frac{l_b^*}{l'_b + l_b^*}\right)^{\sum_{u \in \mathcal{U}} x_{ub}^* w_u} =$$
$$- \sum_{b \in \mathcal{B}} \sum_{u \in \mathcal{U}} x'_{ub} w_u \log\left(\frac{l_b^*/l'_b}{1 + l_b^*/l'_b}\right)^{\frac{\sum_{v \in \mathcal{U}} x_{vb}^* w_v}{\sum_{v \in \mathcal{U}} x'_{vb} w_v}} =$$
$$- \sum_{b \in \mathcal{B}} \sum_{u \in \mathcal{U}} x'_{ub} w_u \log\left(\frac{l_b^*/l'_b}{1 + l_b^*/l'_b}\right)^{l_b^*/l'_b}$$

Given that $(x/(1+x))^x > 1/e$ for $x \geq 0$, we obtain the following bound:

$$W(\mathbf{x}^*, \mathbf{f}^*) - W(\mathbf{x}', \mathbf{f}') \leq \sum_{b \in \mathcal{B}} \sum_{u \in \mathcal{U}} x'_{ub} w_u \log(e) = \log(e)$$

Since $\mathbf{x}^* = \mathbf{x}^{MORA}$ and $\mathbf{f}^* = \mathbf{f}^{MORA}$, this proves the theorem. □

*Proof of Theorem 7:* The proof of the theorem is based on the following steps:
**Step 1**: we first show that while there is some user for which $r_u^{new} \geq e \cdot r_u^{old}$, $W(\mathbf{x}^i, \mathbf{f}^i)$ increases at each iteration until we converge to a region that satisfies $r_u^{new} \leq e \cdot r_u^{old}$ for all $u$.
**Step 2**: we then show that if $r_u^{new} \leq e \cdot r_u^{old} \forall u$, it follows that $W(\mathbf{x}^i, \mathbf{f}^i) \geq W(\mathbf{x}^{MORA}, \mathbf{x}^{MORA}) - 2\log(e)$.
**Step 3**: we further prove that if a subsequent iteration $i$ yields $r_u^{new} \geq e \cdot r_u^{old}$ for some user $u$, then it must be that $W(\mathbf{x}^i, \mathbf{f}^i) \geq W(\mathbf{x}^{MORA}, \mathbf{x}^{MORA}) - (2 + \max_u w_u)\log(e)$.
**Step 4**: finally, we prove that after an iteration such as the above, in the subsequent iterations $W(\mathbf{x}^i, \mathbf{f}^i)$ increases, until we converge once again to a region where $r_u^{new} \leq e \cdot r_u^{old} \forall u$.

We next prove each of the above steps.

**Step 1**: *While there is some user for which $r_u^{new} \geq e \cdot r_u^{old}$, $W(\mathbf{x}^i, \mathbf{f}^i)$ increases at each iteration until we converge to a region that satisfies $r_u^{new} \leq e \cdot r_u^{old}$ for all $u$.*

To prove the above, we consider a variation of the Greedy Largest Gain in which a user only moves to a new location if $r_u^{new} \geq e \cdot r_u^{old}$, and show that this algorithm is guaranteed to converge. To show this, we prove that the network utility function $W(\mathbf{x}, \mathbf{f})$ is a generalized ordinal potential for the algorithm variation. Consider the $i^{th}$ iteration in the algorithm corresponding to a reassociation of user $u$, and let $(\mathbf{x}^{i-1}, \mathbf{f}^{i-1})$ denote the configuration before this iteration and $(\mathbf{x}^i, \mathbf{f}^i)$ the configuration after the iteration. By construction of the algorithm, the following is satisfied:

$$r_u(\mathbf{x}^i, \mathbf{f}^i) \geq e \cdot r_u(\mathbf{x}^{i-1}, \mathbf{f}^{i-1})$$

Let $b$ be the new base station user $u$ associates with, and $a$ her previous base station. Then,

$$W(\mathbf{x}^i, \mathbf{f}^i) - W(\mathbf{x}^{i-1}, \mathbf{f}^{i-1}) =$$
$$\sum_{v \in \mathcal{U}} x_{va}^i w_v \log\left(\frac{\sum_{y \in \mathcal{U}} x_{ya}^i w_y + w_u}{\sum_{y \in \mathcal{U}} x_{ya}^i w_y}\right) +$$
$$\sum_{v \in \mathcal{U} \setminus \{u\}} x_{vb}^i w_v \log\left(\frac{\sum_{y \in \mathcal{U} \setminus \{u\}} x_{yb}^i w_y}{\sum_{y \in \mathcal{U} \setminus \{u\}} x_{yb}^i w_y + w_u}\right) +$$
$$w_u \log(r_u(\mathbf{x}^i, \mathbf{f}^i)) - w_u \log(r_u(\mathbf{x}^{i-1}, \mathbf{f}^{i-1})) =$$
$$l_a^i \log\left(\frac{l_a^i + w_u}{l_a^i}\right) + l_b^{i-1} \log\left(\frac{l_b^{i-1}}{l_b^{i-1} + w_u}\right) +$$
$$w_u \log(r_u(\mathbf{x}^i, \mathbf{f}^i)) - w_u \log(r_u(\mathbf{x}^{i-1}, \mathbf{f}^{i-1}))$$

Since $l_a^i \log\left(\frac{l_a^i + w_u}{l_a^i}\right) \geq 0$, we have

$$W(\mathbf{x}^i, \mathbf{f}^i) - W(\mathbf{x}^{i-1}, \mathbf{f}^{i-1}) \geq$$
$$w_u \log\left(\frac{l_b^{i-1}/w_u}{1 + l_b^{i-1}/w_u}\right)^{\frac{l_b^{i-1}}{w_u}} + w_u \log\left(\frac{r_u(\mathbf{x}^i, \mathbf{f}^i)}{r_u(\mathbf{x}^{i-1}, \mathbf{f}^{i-1})}\right)$$
$$> w_u \log(1/e) + w_u \log(e) = 0 \qquad (14)$$

so that $W(\mathbf{x}, \mathbf{f})$ is a generalized ordinal potential. This implies that the potential game corresponding to the algorithm variation has the finite improvement property; therefore, the algorithm variation converges in a finite number of iterations to a solution that satisfies $r_u^{new} \leq e \cdot r_u^{old} \forall u$. Also, from (14) it follows that $W(\mathbf{x}^i, \mathbf{f}^i) > W(\mathbf{x}^{i-1}, \mathbf{f}^{i-1})$, i.e., the network utility increases at each iteration.

As the Greedy Largest Gain algorithm always selects the user with the largest $r_u^{new}/r_u^{old}$, it will select a user for which $r_u^{new} \geq e \cdot r_u^{old}$, as long as there is one that satisfies this condition, and hence will follow the same steps as the algorithm variation that we have considered above. This implies that there will be some iteration $i$ in which the Greedy Largest Gain algorithm will reach a solution $(\mathbf{x}^i, \mathbf{f}^i)$ that satisfies $r_u^{new} \leq e \cdot r_u^{old}\ \forall u$ and, until reaching this solution, $W(\mathbf{x}^i, \mathbf{f}^i)$ will increase at each iteration.

**Step 2**: *If $r_u^{new} \leq e \cdot r_u^{old}\ \forall u$, it follows that $W(\mathbf{x}^i, \mathbf{f}^i) \geq W(\mathbf{x}^{MORA}, \mathbf{x}^{MORA}) - 2\log(e)$.*

Let $(\mathbf{x}^i, \mathbf{f}^i)$ be the solution at the $i^{th}$ iteration which satisfies $r_u^{new} \leq e \cdot r_u^{old}\ \forall u$. Equation (13) for this solution can be rewritten as

$$\sum_{b \in \mathcal{B}} x_{ub}^i w_u \log\left(\frac{w_u c_{ub}}{\sum_{v \in \mathcal{U}} x_{vb}^i w_v}\right) \geq$$

$$\sum_{b \in \mathcal{B}} x_{ub}^{MORA} w_u \log\left(\frac{w_u c_{ub}}{\sum_{v \in \mathcal{U}} x_{vb}^i w_v + w_u}\right) - w_u log(e)$$

Starting from the above equation and applying the same reasoning as in the proof of Theorem 6 yields $W(\mathbf{x}^i, \mathbf{f}^i) \geq W(\mathbf{x}^{MORA}, \mathbf{f}^{MORA}) - 2\log(e)$.

**Step 3**: *If a subsequent iteration $i$ yields $r_u^{new} \geq e \cdot r_u^{old}$ for some user $u$, then it must be that $W(\mathbf{x}^i, \mathbf{f}^i) \geq W(\mathbf{x}^{MORA}, \mathbf{x}^{MORA}) - (2 + \max_u w_u)\log(e)$.*

Let us that for some iteration $i$ of the algorithm such that it holds $r_u^{new} \leq e \cdot r_u^{old}\ \forall u$ for the solution before this iteration, and $r_u^{new} \leq e \cdot r_u^{old}$, for some $u$, for the solution after the iteration. Let $(\mathbf{x}^{i-1}, \mathbf{f}^{i-1})$ be the solution before iteration $i$ and $(\mathbf{x}^i, \mathbf{f}^i)$ the solution after the iteration. As we have seen above, for the former it holds $W(\mathbf{x}^{i-1}, \mathbf{f}^{i-1}) \geq W(\mathbf{x}^{MORA}, \mathbf{f}^{MORA}) - 2\log(e)$. Let us consider that at iteration $i$ user $u$ moves to base station $b$. Then,

$$W(\mathbf{x}^i, \mathbf{f}^i) - W(\mathbf{x}^{i-1}, \mathbf{f}^{i-1}) \geq$$

$$\sum_{v \in \mathcal{U}} x_{vb}^{i-1} w_v \log\left(\frac{\sum_{t \in \mathcal{U}} x_{tb}^{i-1} w_t}{\sum_{t \in \mathcal{U}} x_{tb}^{i-1} w_t + w_u}\right) =$$

$$w_u \log\left(\frac{\sum_{t \in \mathcal{U}} x_{tb}^{i-1} w_t / w_u}{1 + \sum_{t \in \mathcal{U}} x_{tb}^{i-1} w_t / w_u}\right)^{\frac{\sum_{t \in \mathcal{U}} x_{tb}^{i-1} w_t}{w_u}} \geq$$

$$- w_u log(e) \geq - \max_u w_u log(e)$$

Thus,

$$W(\mathbf{x}^i, \mathbf{f}^i) \geq W(\mathbf{x}^{MORA}, \mathbf{f}^{MORA}) - (2 + \max_u w_u)\log(e).$$

**Step 4**: *After an iteration such as the above, in the subsequent iterations $W(\mathbf{x}^i, \mathbf{f}^i)$ increases, until we converge once again to a region where $r_u^{new} \leq e \cdot r_u^{old}\ \forall u$.*

Let us consider that before iteration $i$ there is some $u$ for which $r_u^{new} \geq e \cdot r_u^{old}$. Then,

$$W(\mathbf{x}^i, \mathbf{f}^i) - W(\mathbf{x}^{i-1}, \mathbf{f}^{i-1}) \geq w_u \log(r^{new}) - w_u \log(r^{old}) +$$

$$\sum_{v \in \mathcal{U}} x_{vb}^{i-1} w_v \log\left(\frac{\sum_{t \in \mathcal{U}} x_{tb}^{i-1} w_t}{\sum_{t \in \mathcal{U}} x_{tb}^{i-1} w_t + w_u}\right) >$$

$$w_u \log(e) - w_u \log(e) \geq 0$$

Therefore, if at some iteration we get $r_u^{new} \geq e \cdot r_u^{old}$ for some $u$, then for that iteration it will hold $W(\mathbf{x}^i, \mathbf{f}^i) \geq W(\mathbf{x}^{MORA}, \mathbf{f}^{MORA}) - (2 + \max_u w_u)\log(e)$, and from this point on $W(\mathbf{x}^i, \mathbf{f}^i)$ is going to increase until we reach $W(\mathbf{x}^i, \mathbf{f}^i) \geq W(\mathbf{x}^{MORA}, \mathbf{f}^{MORA}) - 2\log(e)$ again. □